\documentclass{article}
\usepackage{ijcai19}

\usepackage{xcolor}
\usepackage{times}
\usepackage{soul}
\usepackage{url}
\usepackage[hidelinks]{hyperref}
\usepackage[T1]{fontenc}
\usepackage{textcomp}
\usepackage[utf8]{inputenc}
\usepackage[small]{caption}
\usepackage[subrefformat=parens,labelformat=parens]{subfig}
\usepackage{graphicx}
\usepackage{amsfonts}
\usepackage{amssymb}
\usepackage{amsthm}
\usepackage{amsmath}
\usepackage{makecell}
\usepackage{booktabs}
\usepackage[linesnumbered,ruled,vlined]{algorithm2e}
\SetKwRepeat{Do}{do}{while}
\DontPrintSemicolon

\SetCommentSty{comment}

\usepackage{latexsym}

\title{Hierarchical Graph Convolutional Networks \\ for Semi-supervised Node Classification}

\author{
Fenyu Hu$^{1,2}$\footnote{The first two authors contributed equally to this work.}\and
Yanqiao Zhu$^{1,2}$\footnotemark[1]\and
Shu Wu$^{1,2}$\footnote{To whom correspondence should be addressed.}\and
Liang Wang$^{1,2}$\And
Tieniu Tan$^{1,2}$\\
\affiliations
$^1$
University of Chinese Academy of Sciences\\
$^2$
Center for Research on Intelligent Perception and Computing, National Laboratory of Pattern Recognition, Institute of Automation, Chinese Academy of Sciences\\
\emails
\{fenyu.hu,yanqiao.zhu\}@cripac.ia.ac.cn,
\{shu.wu,wangliang,tnt\}@nlpr.ia.ac.cn
}

\begin{document}

\maketitle

\begin{abstract}
Graph convolutional networks (GCNs) have been successfully applied in node classification tasks of network mining. However, most of these models based on neighborhood aggregation are usually shallow and lack the ``graph pooling'' mechanism, which prevents the model from obtaining adequate global information. In order to increase the receptive field, we propose a novel deep Hierarchical Graph Convolutional Network (H-GCN) for semi-supervised node classification. H-GCN first repeatedly aggregates structurally similar nodes to hyper-nodes and then refines the coarsened graph to the original to restore the representation for each node. Instead of merely aggregating one- or two-hop neighborhood information, the proposed coarsening procedure enlarges the receptive field for each node, hence more global information can be captured. The proposed H-GCN model shows strong empirical performance on various public benchmark graph datasets, outperforming state-of-the-art methods and acquiring up to 5.9\% performance improvement in terms of accuracy. In addition, when only a few labeled samples are provided, our model gains substantial improvements.
\end{abstract}

\section{Introduction}

Graphs nowadays become ubiquitous owing to the ability to model complex systems such as social relationships, biological molecules, and publication citations. The problem of classifying graph-structured data is fundamental in many areas. Besides, since there is a tremendous amount of unlabeled data in nature and labeling data is often expensive and time-consuming, it is often challenging and crucial to analyze graphs in a semi-supervised manner. For instance, for semi-supervised node classification in citation networks, where nodes denote articles and edges represent citation, the task is to predict the label of every article with only a few labeled data.

As an efficient and effective approach to graph analysis, network embedding has attracted a lot of research interests. It aims to learn low-dimensional representations for nodes whilst still preserving the topological structure and node feature attributes. Many methods have been proposed for network embedding, which can be used in the node classification task, such as DeepWalk \cite{Perozzi:2014:DOL:2623330.2623732} and node2vec \cite{Grover:2016:NSF:2939672.2939754}. They convert the graph structure into sequences by performing random walks on the graph. Then, the proximity between the nodes can be captured based on the co-occurrence statistics in these sequences. But they are unsupervised algorithms and cannot perform node classification tasks in an end-to-end fashion. Unlike previous random-walk-based approaches, employing neural networks on graphs has been studied extensively in recent years. Using an information diffusion mechanism, the graph neural network (GNN) model updates states of the nodes and propagate them until a stable equilibrium \cite{Scarselli:2009ku}. Both of the highly non-linear topological structure and node attributes are fed into the GNN model to obtain the graph embedding. Recently, there is an increasing research interest in applying convolutional operations on the graph. These graph convolutional networks (GCNs) \cite{Kipf:2016tc,Velickovic:2018we} are based on the neighborhood aggregation scheme which generates node embedding by combining information from neighborhoods. Comparing with conventional methods, GCNs achieve promising performance in various graph analytical tasks such as node classification and graph classification \cite{NIPS2016_6081} and has shown effective for many application domains, for instance, recommendation \cite{Wu:2019vb,Cui:2019:DWO:3308558.3313444}, traffic forecasting \cite{Yu:2018fqba}, and action recognition \cite{stgcn_aaai18}.

Nevertheless, GCN-based models are usually shallow and lack the ``graph pooling'' mechanism, which restricts the scale of the receptive field. For example, there are only two layers in GCN \cite{Kipf:2016tc}. As each graph convolutional layer acts as the approximation of aggregation on the first-order neighbors, the two-layer GCN model only aggregates information from two-hop neighborhoods for each node. Because of the restricted receptive field, the model has difficulty in obtaining adequate global information. However, it has been observed from the reported results \cite{Kipf:2016tc} that simply adding more layers will degrade the performance. As explained in \cite{AAAI1816098}, each GCN layer acts as a form of Laplacian smoothing in essence, which makes the features of nodes in the same connected component similar. Thereby, adding too many convolutional layers will result in the output features over-smoothed and make them indistinguishable. Meanwhile, deeper neural networks with more parameters are harder to train. Although some recent methods \cite{AAAI1816273,pmlr-v80-xu18c,Ying2018HierarchicalGR} try to get the global information through deeper models, they are either unsupervised models or need many training examples. As a result, they are still not capable of solving the semi-supervised node classification task directly.

\begin{figure}
	\centering
	\includegraphics[width=\columnwidth]{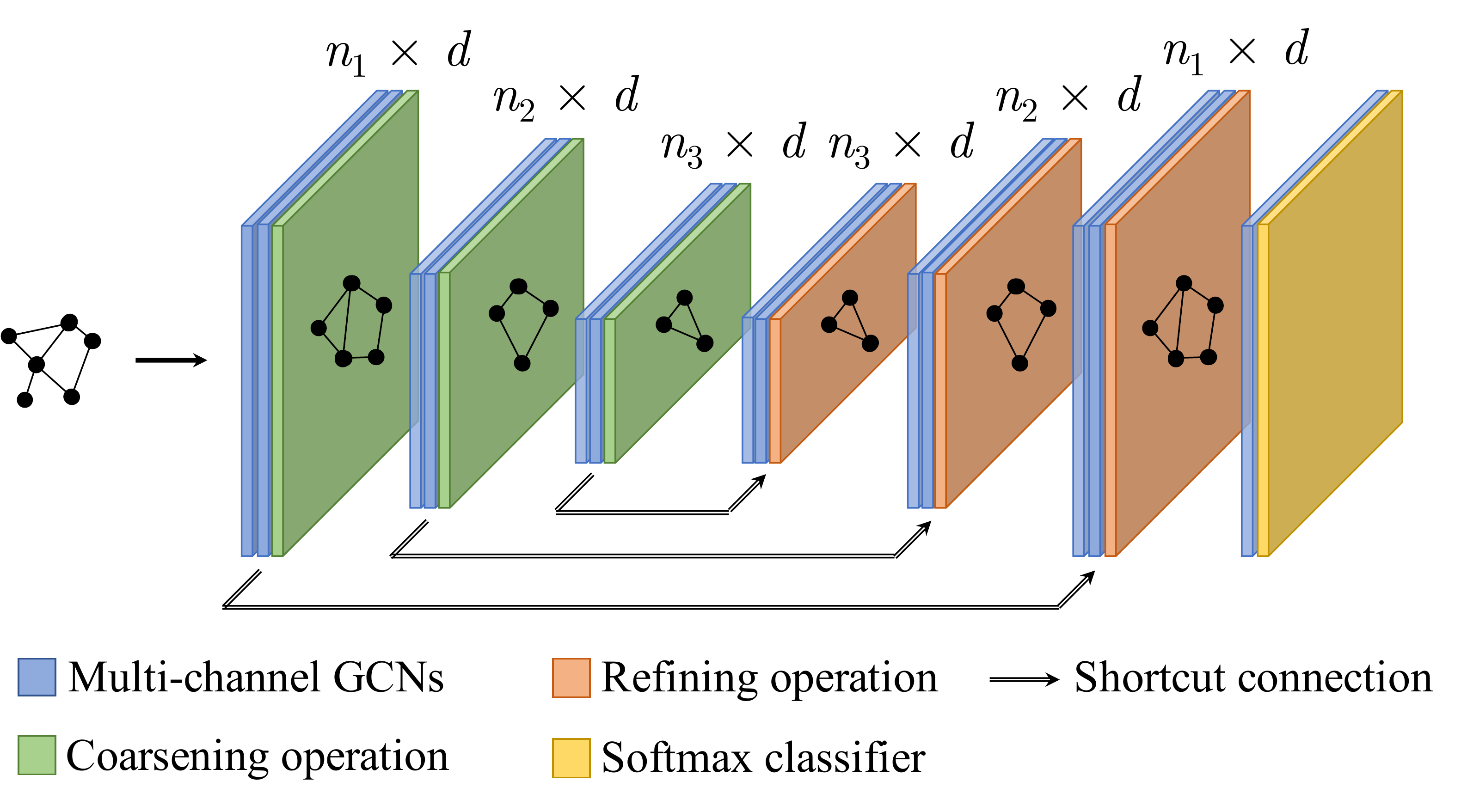}
	\caption{The workflow of H-GCN. In this illustration, there are seven layers with three coarsening layers, three symmetric refining layers, and one output layer. Coarsening layer at level $i$ produces graph $\mathcal{G}_{i+1}$ of $n_{i+1}$ hyper-nodes with $d$-dimensional latent representations, vice versa for refining layers.}
	\label{fig:workflow}
\end{figure}

To this end, we propose a novel architecture of \underline{H}ierarchical \underline{G}raph \underline{C}onvolutional \underline{N}etworks, H-GCN for brevity, for node classification on graphs\footnote{To make our results reproducible, all relevant source codes are publicly available at \url{https://github.com/CRIPAC-DIG/H-GCN}.}. Inspired from the flourish of applying deep architectures and the pooling mechanism into image classification tasks, we design a deep hierarchical model with coarsening mechanisms. The H-GCN model increases the receptive field of graph convolutions and can better capture global information. As illustrated in Figure \ref{fig:workflow}, H-GCN mainly consists of several coarsening layers and refining layers. For each coarsening layer, the graph convolutional operation is first conducted to learn node representations. Then, a coarsening operation is performed to aggregate structurally similar nodes into hyper-nodes. After the coarsening operation, each hyper-node represents a local structure of the original graph, which can facilitate exploiting global structures on the graph. Following coarsening layers, we apply symmetric graph refining layers to restore the original graph structure for node classification tasks. Such a hierarchical model manages to comprehensively capture nodes' information from local to global perspectives, leading to better node representations.

The main contributions of this paper are twofold. Firstly, to the best of our knowledge, it is the first work to design a deep hierarchical model for the semi-supervised node classification task. Compared to previous work, the proposed model consists of more layers with larger receptive fields, which is able to obtain more global information through the coarsening and refining procedures. Secondly, we conduct extensive experiments on a variety of public datasets and show that the proposed method constantly outperforms other state-of-the-art approaches. Notably, our model gains a considerable improvement over other approaches with very few labeled samples provided for each class.


\section{Related Work}

In this section, we review some previous work on graph convolutional networks for semi-supervised node classification, hierarchical representation learning on graphs, and graph reduction algorithms.

\subsubsection{Graph Convolutional Networks}
In the past few years, there has been a surge of applying convolutions on graphs. These approaches are essentially based on the neighborhood aggregation scheme and can be further divided into two branches: spectral approaches and spatial approaches.

The spectral approaches are based on the spectral graph theory to define parameterized filters. \citeauthor{Bruna:2014vg} (\citeyear{Bruna:2014vg}) first define the convolutional operation in the Fourier domain. However, its heavy computational burden limits the application to large-scale graphs. In order to improve efficiency, \citeauthor{NIPS2016_6081} (\citeyear{NIPS2016_6081}) propose ChebNet to approximate the $K$-polynomial filters by means of a Chebyshev expansion of the graph Laplacian. \citeauthor{Kipf:2016tc} (\citeyear{Kipf:2016tc}) further simplify the ChebNet by truncating the Chebyshev polynomial to the first-order neighborhood. DGCN \cite{Zhuang:2018:DGC:3178876.3186116} uses random walks to construct a positive mutual information matrix. Then, it utilizes that matrix along with the graph's adjacency matrix to encode both local consistency and global consistency.

The spatial approaches generate node embedding by combining the neighborhood information in the vertex domain. MoNet \cite{8100059} and SplineCNN \cite{Fey/etal/2018} integrate the local signals by designing a universe patch operator. To generalize to unseen nodes in an inductive setting, GraphSAGE \cite{DBLP:conf/nips/HamiltonYL17} samples a fixed number of neighbors and employs several aggregation functions, such as concatenation, max-pooling, and LSTM aggregator. GAT \cite{Velickovic:2018we} introduces the attention mechanism to model different influences of neighbors with learnable parameters. \citeauthor{Gao:2018:LLG:3219819.3219947} (\citeyear{Gao:2018:LLG:3219819.3219947}) select a fixed number of neighborhood nodes for each feature and enables the use of regular convolutional operations on Euclidean spaces. However, the above two branches of GCNs are usually shallow and cannot obtain adequate global information as a consequence.

\subsubsection{Hierarchical Representation Learning on Graphs}
Some work has been proposed for learning hierarchical information on graphs. \citeauthor{AAAI1816273} (\citeyear{AAAI1816273}) and \citeauthor{2018arXiv180209612L} (\citeyear{2018arXiv180209612L}) use a coarsening procedure to construct a coarsened graph of smaller size and then employ unsupervised methods, such as DeepWalk \cite{Perozzi:2014:DOL:2623330.2623732} and node2vec \cite{Grover:2016:NSF:2939672.2939754} to learn node embedding based on that coarsened graph. Then, they conduct a refining procedure to get the original graph embedding. However, their two-stage methods are not capable of utilizing node attribute information and can neither perform node classification task in an end-to-end fashion. JK-Nets \cite{pmlr-v80-xu18c} proposes general layer aggregation mechanisms to combine the output representation in every GCN layer. However, it can only propagate information across edges of the graph and are unable to aggregate information hierarchically. Therefore, the hierarchical structure of the graph cannot be learned by JK-Nets. To solve this problem, DiffPool \cite{Ying2018HierarchicalGR} proposes a pooling layer for graph embedding to reduce the size by a differentiable network. As DiffPool is designed for graph classification tasks, it cannot generate embedding for every node in the graph; hence it cannot be directly applied in node classification scenarios.

\subsubsection{Graph Reduction}
Many approaches have been proposed to reduce the graph size without losing too much information, which facilitate downstream network analysis tasks such as community discovery and data summarization. There are two main classes of methods that reduce the graph size: graph sampling and graph coarsening. The first category is based on graph sampling strategy \cite{6104045,DBLP:journals/corr/HuL13,8000122}, which might lose key information during the sampling process. The second category applies graph coarsening strategies that collapse structure-similar nodes into hyper-nodes to generate a series of increasingly coarser graphs. The coarsening operation typically consists of two steps, i.e. grouping and collapsing. At first, every node is assigned to groups in a heuristic manner. Here a group refers to a set of nodes that constitute a hyper-node. Then, these groups are used to generate a coarser graph. For an unmatched node, \citeauthor{Hendrickson:1995:MAP:224170.224228} (\citeyear{Hendrickson:1995:MAP:224170.224228}) randomly select one of its un-matched neighbors and merge these two nodes. \citeauthor{doi:10.1137/S1064827595287997} (\citeyear{doi:10.1137/S1064827595287997}) merge the two un-matched nodes by selecting those with the maximum weight edge. \citeauthor{LaSalle:2015:MMB:2780684.2780857} (\citeyear{LaSalle:2015:MMB:2780684.2780857}) use a secondary jump during grouping.

However, these graph reduction approaches are usually used in unsupervised scenarios, such as community detection and graph partition. For semi-supervised node classification tasks, existing graph reduction methods cannot be used directly, as they are not capable of learning complex attributive and structural features of graphs. In this paper, H-GCN conducts graph reduction for non-Euclidean geometry like the pooling mechanism for Euclidean data. In this sense, our work bridges graph reduction for unsupervised tasks to the practical but more challenging semi-supervised node classification problems.

\section{The Proposed Method}

\subsection{Preliminaries}

\subsubsection{Notations and Problem Definition}

For the input undirected graph $\mathcal{G}_1 = (\mathcal{V}_1, \mathcal{E}_1)$, where $\mathcal{V}_1$ and $\mathcal{E}_1$ are respectively the set of $n_1$ nodes and $e_1$ edges, let $A_1 \in \mathbb{R}^{n_1 \times n_1}$ be the adjacency matrix describing its edge weights and $X \in \mathbb{R}^{n_1 \times d_1}$ be the node feature matrix, where $d_1$ is the dimension of the attributive features. We use edge weights to indicate connection strengths between nodes. For the H-GCN network, the graph fed into the $i$\textsuperscript{th} layer is represented as $\mathcal{G}_i$ with $n_i$ nodes. The adjacency matrix and hidden representation matrix of $\mathcal{G}_i$ are represented by $A_i \in \mathbb{R}^{n_i \times n_i}$ and $H_i \in \mathbb{R}^{n_i \times d_i}$ respectively.

Since coarsening layers and refining layers are symmetrical, $A_i$ is identical to $A_{l - i + 1}$, where $l$ is the total number of layers in the network. For example, in the seven-layer model illustrated in Figure \ref{fig:workflow}, $\mathcal{G}_3$ is the input graph for the third layer and $\mathcal{G}_5$ is the resulting graph from the fourth layer. After one coarsening operation and one refining operation, $\mathcal{G}_3$ and $\mathcal{G}_5$ share exactly the same topological structure $A_3$. As nodes will be assigned as a hyper-node, we define {\it node weight} as the number of nodes contained in a hyper-node.

Given the labeled node set $\mathcal{V}_L$ containing $m \ll n_1$ nodes, where each node $v_i \in \mathcal{V}_L$ is associated with a label $y_i \in \mathcal{Y}$, our objective is to predict labels of $\mathcal{V} \backslash \mathcal{V}_L$.

\subsubsection{Graph Convolutional Networks}

Graph convolutional networks achieve promising generalization in various tasks and our work is built upon the GCN module. At layer $i$, taking graph adjacency matrix $A_{i}$ and hidden representation matrix $H_{i}$ as input, each GCN module outputs a hidden representation matrix $G_i \in \mathbb{R}^{n_i \times d_{i+1}}$, which is described as:
\begin{equation}
	G_i = \operatorname{ReLU} \left( \tilde{D}^{-\frac {1}{2}}_{i} \tilde{A}_{i} \tilde{D}^{-\frac{1}{2}}_{i} H_{i} \theta_i \right),
	\label{eq:GCN}
\end{equation}
where $H_1 = X$, $\operatorname{ReLU}(x) = \max(0,x)$, adjacency matrix with self-loop $\tilde{A}_{i} = A_{i} + I$,  $\tilde{D}_{i}$ is the degree matrix of $\tilde{A}_{i}$, and $\theta_i \in \mathbb{R}^{d_{i} \times d_{i + 1}}$ is a trainable weight matrix. For ease of parameter tuning, we set output dimension $d_i = d$ for all coarsening and refining layers throughout this paper.

\subsection{The Overall Architecture}

For a H-GCN network of $l$ layers, the $i$\textsuperscript{th} graph coarsening layer first conducts a graph convolutional operation as formulated in Eq. (\ref{eq:GCN}) and then aggregates structurally similar nodes into hyper-nodes, producing a coarser graph $\mathcal{G}_{i+1}$ and node embedding matrix $H_{i+1}$ with fewer nodes. The corresponding adjacent matrix $A_{i+1}$ and $H_{i+1}$ will be fed into the $(i + 1)$\textsuperscript{th} layer. Symmetrically, the graph refining layer also performs a graph convolution at first and then refines the coarsened graph to restore the finer graph structure. In order to boost optimization in deeper networks, we add shortcut connections \cite{7780459} across each coarsened graph and its corresponding refined part.

Since the topological structure of the graph changes between layers, we further introduce a node weight embedding matrix $S_i$, which transforms the number of nodes contained in each hyper-node into real-valued vectors. Both of the node weight embedding and $H_i$ will be fed into the $i$\textsuperscript{th} layer. Besides, we add multiple channels by employing different GCNs to explore different feature subspaces.

The graph coarsening layers and refining layers altogether integrate different levels of node features and thus avoid over-smoothing during repeated neighborhood aggregation. After the refining process, we obtain a node embedding matrix $H_{l - 1}\in \mathbb{R}^{n_1 \times d}$, where each row represents a node representation vector. In order to classify each node, we apply an additional GCN module followed by a softmax classifier on $H_{l - 1}$.

\subsection{The Graph Coarsening Layer}

Every graph coarsening layer consists of two steps, i.e. graph convolution and graph coarsening. A GCN module is firstly used to extract structural and attributive features by aggregating neighborhoods' information as described in Eq. (\ref{eq:GCN}). For the graph coarsening procedure, we design the following two hybrid grouping strategies to assign nodes with similar structures into a hyper-node in the coarser graph. We first conduct structural equivalence grouping, followed by structural similarity grouping.

\begin{figure}
	\centering
	\includegraphics[width=\columnwidth]{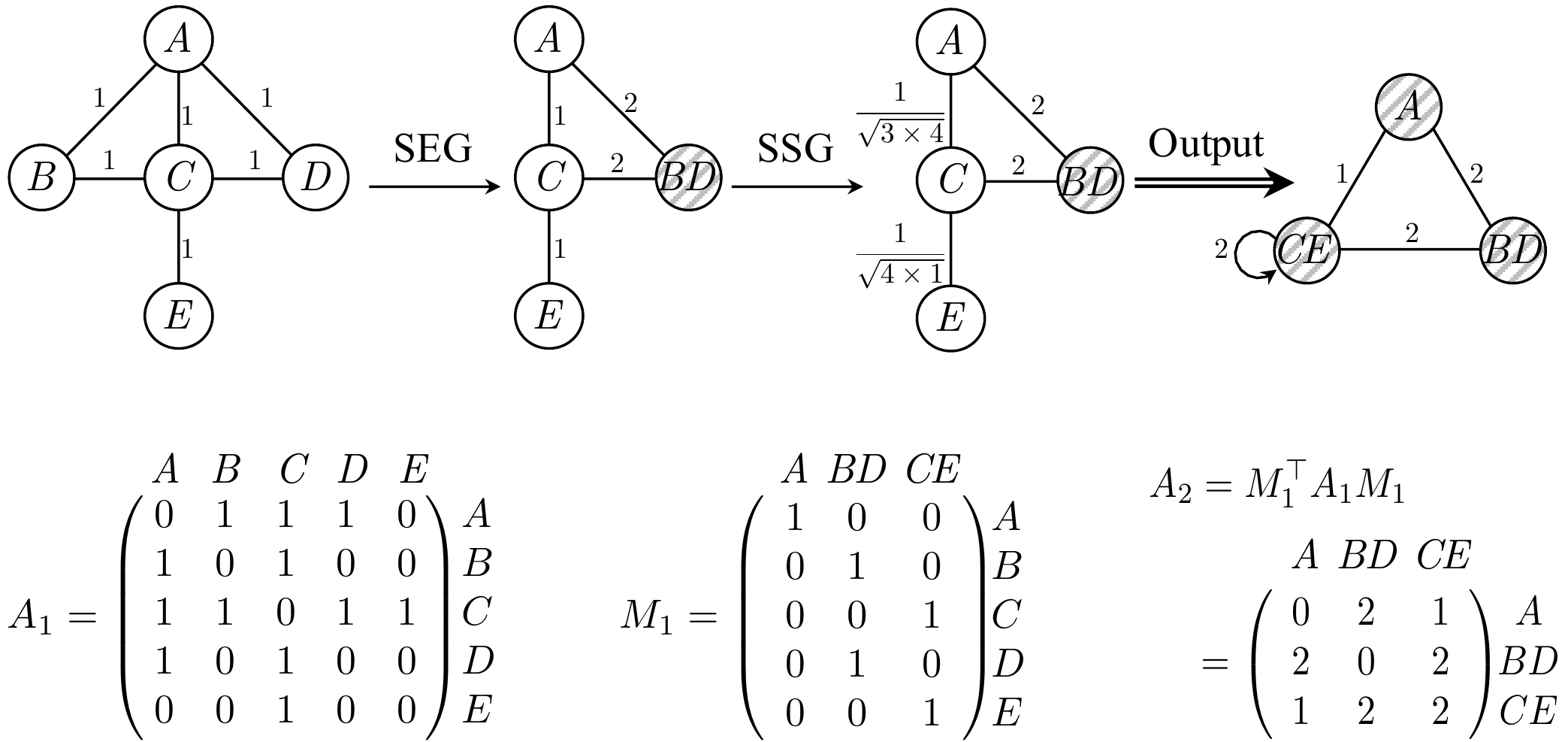}
	\caption{The graph coarsening operation of a toy graph. Numbers indicate edge weights and nodes in shadow are hyper-nodes. In SEG, node $B$ and $D$ share the same neighbors, so they are grouped into a hyper-node. In SSG, node $C$ and $E$ are grouped because they have the largest normalized connection weight. Node $A$ constitutes a hyper-node by itself since it remains unmarked.}
	\label{fig:coarsening}
\end{figure}

\paragraph{Structural equivalence grouping (SEG).}
If two nodes share the same set of neighbors, they are considered to be structurally equivalent. We then assign these two nodes to be a hyper-node. For example, as illustrated in Figure \ref{fig:coarsening}, nodes $B$ and $D$ are structurally equivalent, so these two nodes are allocated as a hyper-node. We mark all these structurally equivalent nodes and leave other nodes unmarked to avoid repetitive grouping operation on nodes.

\paragraph{Structural similarity grouping (SSG).}
Then, we calculate the {\it structural similarity} between the unmarked node pairs $(v_j, v_k)$ as the normalized connection strength $s(v_j, v_k)$:
\begin{equation}
	s(v_j, v_k) = \frac{A_{jk}}{\sqrt{D(v_j) \cdot D(v_k)}},
	\label{eq:normalized-connection-strength}
\end{equation}
where $A_{jk}$ is the edge weight between $v_j$ and $v_k$, and $D(\cdot)$ is the node weight.

We iteratively take out an unmarked node $v_j$ and calculate normalized connection strengths with all its unmarked neighbors. After that, we select its neighbor node $v_k$ which has the largest structural similarity to form a new hyper-node and mark the two nodes. Particularly, if one node is left unmarked and all of its immediate neighbors are marked, it will be marked as well and constitutes a hyper-node by itself. For example, in Figure \ref{fig:coarsening}, node pair $(C, E)$ has the largest structural similarity, so they are grouped together to form a hyper-node. After that, since only node $A$ remains unmarked, it constitutes a hyper-node by itself.

Please note that if we take out unmarked nodes in a different order, the resulting hyper-graph will be different. The later we take a node out, the less its neighbors will be left unmarked. So, for a node with fewer neighbors, it has fewer probabilities to be grouped when it is taken out late. Therefore, we take out the unmarked nodes in ascending order according to the number of neighbors.

Using the above two grouping strategies, we are able to acquire all the hyper-nodes. For one hyper-node $v_i$, its edge weight to $v_j$ is the summation over edge weights of $v_j$'s neighbor nodes contained in $v_i$. The updated node weights and edge weights will be used in Eq. (\ref{eq:normalized-connection-strength}) in the next coarsening layer.

In order to help restore the coarsened graph to original graph, we preserve the grouping relationship between nodes and their corresponding hyper-nodes in a matrix $M_i \in \mathbb{R}^{n_i \times n_{i + 1}}$. Formally, at layer $i$, entry $m_{jk}$ in the grouping matrix $M_i$ is calculated as:
\begin{equation}
	m_{jk} = \left\{
		\begin{array}{ll}
			1, & \mbox{if $v_j$ in $\mathcal{G}_i$ is grouped into $v_k$ in $\mathcal{G}_{i + 1}$};\\
        	0, & \mbox{otherwise}.
        \end{array}
    \right.
    \label{eq:matching-matrix}
\end{equation}
An example of the coarsening operation on a toy graph is given in Figure \ref{fig:coarsening}. Note that $m_{11} = 1$ in this illustration, since node $A$ constitutes its hyper-node by itself. Next, the hidden node embedding matrix is determined as:
\begin{equation}
	H_{i + 1} = M_i^\top \cdot G_i.
	\label{eq:node-representation}
\end{equation}
In the end, we generate a coarser graph $\mathcal{G}_{i+1}$, whose adjacency matrix can be calculated as:
\begin{equation}
	A_{i+1} = M_i^\top \cdot A_{i} \cdot M_i.
	\label{eq:coarsened-graph}
\end{equation}

The coarser graph $\mathcal{G}_{i+1}$ along with the resulting representation matrix $H_{i+1}$ will be fed into the next layer as input. The resulting node embedding to generate in each coarsening layer will then be of lower resolution. The graph coarsening procedure is summarized in Algorithm \ref{algo:graph-coarsening-layer}.

\begin{algorithm}
	\caption{The graph coarsening operation}
	\label{algo:graph-coarsening-layer}
	\KwIn{Graph $\mathcal{G}_i$ and node representation $H_i$}
	\KwOut{Coarsened graph $\mathcal{G}_{i+1}$ and node representation $H_{i+1}$}
	
	Calculate GCN output $G_i$ according to Eq. (\ref{eq:GCN})\;
	Initialize all nodes as unmarked\;
	
	\tcc{Structural equivalence grouping}
	Group and mark node pairs having the same neighbors\;
	
	\tcc{Structural similarity grouping}
	Sort all unmarked nodes in ascending order according to the number of neighbors\;
	\Repeat{all nodes are marked} {
		\For{each unmarked node $v_j$} {
			\For{each unmark node $v_k$ adjacent to $v_j$} {
				Calculate $s(v_j, v_k)$ according to Eq. (\ref{eq:normalized-connection-strength})\;
			}
			Group and mark the node pair $(v_j, v_k)$ having the largest $s(v_j, v_k)$\;
		}
	}
	
	Update node weights and edge weights\;
	
	Construct grouping matrix $M_i$ according to Eq. (\ref{eq:matching-matrix})\;
	
	Calculate node representation $H_{i+1}$ according to Eq. (\ref{eq:node-representation})\;
	Construct coarsened graph $\mathcal{G}_{i+1}$ according to Eq. (\ref{eq:coarsened-graph})\;
	\Return{$\mathcal{G}_{i+1}, H_{i+1}$}
\end{algorithm}

\subsection{The Graph Refining Layer}

To restore the original topological structure of the graph and further facilitate node classification, we stack the same numbers of graph refining layers as coarsening layers. Like the coarsening procedure, each refining layer contains two steps, namely generating node embedding vectors and restoring node representations.

To learn a hierarchical representation of nodes, a GCN is employed at first. Since we have saved the grouping relationship in the grouping matrix during the coarsening process, we utilize $M_{l - i}$ to restore the refined node representation matrix of layer $i$. We further employ residual connections between the two corresponding coarsening and refining layers. In summary, node representations are computed by:
\begin{equation}
	H_i = M_{l - i} \cdot G_i + G_{l - i}.
\end{equation}

\subsection{Node Weight Embedding and Multiple Channels}

As depicted in Figure \ref{fig:coarsening}, since different hyper-nodes may have different node weights, we assume such consequent node weights could reflect the hierarchical characteristics of coarsened graphs. In order to better capture the hierarchical information, we use the node weight embedding to supplement information in $H_i$. Here we transform the node weight into real-valued vectors by looking up one randomly initialized node weight embedding matrix $V \in \mathbb{R}^{|T|\times p}$, where $T$ is the set of node weights and $p$ is the dimension of the embedding. We apply node weight embedding in every coarsening and refining layer. For graph $\mathcal{G}_i$, we obtain its node weight embedding $S_i \in \mathbb{R}^{n_i \times p}$ by looking up $V$ according to the node weight. For example, if one hyper-node contains three nodes, the third row of $V$ will be selected as its node weight embedding. We then concatenate $H_i$ and $S_i$ and the resulting $(d+p)$-dimensional matrix will be fed into the next GCN layer subsequently.

Multi-channel mechanisms help explore features in different subspaces and H-GCN employs multiple channels on GCN to obtain rich information jointly at each layer. After obtained $c$ channels $\left[ G_{i}^{1}, G_{i}^{2}, \dots, G_{i}^{c} \right]$, we perform weighted average on these feature maps:
\begin{equation}
	G_i = \sum_{j=1}^{c} w_j \cdot G_i^j,
\end{equation}
where $w_j$ is a trainable weight of channel $j$.

\subsection{The Output Layer}

Finally, in the output layer $l$, we use a GCN with a softmax classifier on $H_{l - 1}$ to output probabilities of nodes:
\begin{equation}
	H_l = \operatorname{softmax}\left( \operatorname{ReLU} \left( \tilde{D}^{-\frac {1}{2}}_{l} \tilde{A}_{l} \tilde{D}^{-\frac{1}{2}}_{l} H_{l - 1} \theta_{l} \right) \right),
\end{equation}
where $\theta_{l} \in \mathbb{R}^{d \times |\mathcal{Y}|}$ is a trainable weight matrix and $H_l \in \mathbb{R}^{n_1 \times |\mathcal{Y}|}$ denotes the probabilities of nodes belonging to each class $y \in \mathcal{Y}$.

The loss function is defined as the cross-entropy of predictions over the labeled nodes:
\begin{equation}
	\mathcal{L} = -\sum_{i = 1}^{m}\sum_{y = 1}^{|\mathcal{Y}|} \mathbb{I}(h_i = y_i) \log{P(h_i, y_i)},
\end{equation}
where $\mathbb{I}(\cdot)$ is the indicator function, $y_i$ is the true label for $v_i$, $h_i$ is the prediction for labeled node $v_i$, and $P(h_i, y_i)$ is the predicted probability that $v_i$ is of class $y_i$.

\subsection{Complexity Analysis and Model Comparison}

In this section, we analyze the model complexity and compare it with mainstream graph convolutional models, such as GCN and GAT.

For GCN, preprocessing matrices $\tilde{D}_i^{-\frac{1}{2}} \tilde{A}_i \tilde{D}_i^{-\frac {1}{2}}$ takes $O(n^3)$, and the training process for each layer takes $O(|\mathcal{E}|CF)$, where $\mathcal{E}$ is the edge set and $C, F$ are embedding dimensions. For GAT, the masked attention over all nodes requires $O(n^2)$ in the training process.

For H-GCN, the preprocessing takes $O(n\log n)$ to sort the unmarked nodes and $O(mn)$ for SSG, where $m$ is the average number of neighborhoods. For training, the complexity is also $O(|\mathcal{E}|CF)$. Therefore, H-GCN is as asymptotically efficient as GCN and is more efficient than GAT.

\section{Experiments and Analysis}


\subsection{Experimental Settings}

\subsubsection{Datasets}

For a comprehensive comparison with state-of-the-art methods, we use four widely-used datasets including three citation networks and one knowledge graph. We conduct semi-supervised node classification task in the transductive setting. The statistics of these datasets are summarized in Table \ref{tab:datasets}. We set the node weight and edge weight of the graph to one for all four datasets. The dataset configuration follows the same setting in \cite{pmlr-v48-yanga16,Kipf:2016tc} for a fair comparison. For citation networks, documents and citations are treated as nodes and edges, respectively. For the knowledge graph, each triplet $(e_1, r, e_2)$ will be assigned with separate relation nodes $r_1$ and $r_2$ as $(e_1, r_1)$ and $(e_2, r_2)$, where $e_1$ and $e_2$ are entities and $r$ is the relation between them. During training, only 20 labels per class are used for each citation network and only one label per class is used for NELL during training. Besides, 500 nodes in each dataset are selected randomly as the validation set. We do not use the labels of the validation set for model training.

\begin{table}
	\centering
	\resizebox{\columnwidth}{!}{
	\begin{tabular}{lcccc}
		\toprule
		Dataset & Cora & Citeseer & Pubmed & NELL \\ \midrule
		Type  & \multicolumn{3}{c}{Citation network} & \multicolumn{1}{c}{Knowledge graph} \\
		\# Vertices & 2,708 & 3,327 & 19,717 & 65,755 \\
		\# Edges & 5,429  & 4,732  & 44,338 & 266,144 \\
		\# Classes  & 7   & 6      & 3     & 210 \\
		\# Features & 1,433  & 3,703   & 500   & 5,414 \\
		Labeling rate & 0.052 & 0.036 & 0.003 & 0.003 \\
		\bottomrule
		\end{tabular}
	}
	\caption{Statistics of datasets used in experiments}
	\label{tab:datasets}
\end{table}

\subsubsection{Baseline Methods}

To evaluate the performance of H-GCN, we compare our method with the following representative methods:

\begin{itemize}
	\item {\bf DeepWalk} \cite{Perozzi:2014:DOL:2623330.2623732} generates the node embedding via random walks in an unsupervised manner, then nodes are classified by feeding the embedding vectors into an SVM classifier.
	\item {\bf Planetoid} \cite{pmlr-v48-yanga16} not only learns node embedding but also predicts the context in graph. It also leverages label information to build both transductive and inductive formulations.
	\item {\bf GCN} \cite{Kipf:2016tc} produces node embedding vectors by truncating the Chebyshev polynomial to the first-order neighborhoods.
	\item {\bf GAT} \cite{Velickovic:2018we} generates node embedding vectors by modeling the differences between the node and its one-hop neighbors.
	\item {\bf DGCN} \cite{Zhuang:2018:DGC:3178876.3186116} utilizes the graph adjacency matrix and the positive mutual information matrix to encode both local consistency and global consistency.
\end{itemize}

\subsubsection{Parameter Settings}

We train our model using Adam optimizer with a learning rate of $0.03$ for $250$ epochs. The dropout is applied to all feature vectors with rates of $0.85$. Besides, the $\ell_2$ regularization factor is set to $0.0007$. Considering different scales of datasets, we set the total number of layers $l$ to $9$ for citation networks and $11$ for the knowledge graph, and apply four-channel GCNs in both coarsening and refining layers.

\subsection{Node Classification Results}

To demonstrate the overall performance of semi-supervised node classification, we compare the proposed method with other state-of-the-art methods. The performance in terms of accuracy is shown in Table \ref{tab:results}. The best performance of each column is highlighted in boldface. The performance of our proposed method is reported based on the average of 20 measurements. Note that running GAT on the NELL dataset requires more than 64G memory; hence its performance on NELL is not reported.

\begin{table}
	\centering
	\resizebox{\columnwidth}{!}{
    \begin{tabular}{ccccc}
		\toprule
		Method & Cora  & Citeseer & Pubmed & NELL \\ \midrule
		DeepWalk & 67.2\% & 43.2\% & 65.3\% & 58.1\% \\
		Planetoid & 75.7\% & 64.7\% & 77.2\% & 61.9\% \\
		GCN   & 81.5\% & 70.3\% & 79.0\% & 73.0\% \\
		GAT   & 83.0 \textpm{} 0.7\% & 72.5 \textpm{} 0.7\% & 79.0 \textpm{} 0.3\% & -- \\
		DGCN  & 83.5\% & 72.6\% & 79.3\% & 74.2\% \\
		H-GCN & {\bf 84.5 \textpm{} 0.5\%} & {\bf 72.8 \textpm{} 0.5\%} & {\bf 79.8 \textpm{} 0.4\%} & {\bf 80.1 \textpm{} 0.4\%} \\
	\bottomrule
    \end{tabular}
    }
	\caption{Results of node classification in terms of accuracy}
	\label{tab:results}
\end{table}

The results show that the proposed method consistently outperforms other state-of-the-art methods, which verify the effectiveness of the proposed coarsening and refining mechanisms. Notably, compared with citation networks, H-GCN surpasses other baselines by larger margins on the NELL dataset. To be specific, the accuracy of H-GCN exceeds GCN and DGCN by 7.1\% and 5.9\% on NELL dataset respectively. We analyze the results as follows.

Regarding traditional random-walk-based algorithms such as DeepWalk and Planetoid, their performance is relatively poor. DeepWalk cannot model the attribute information, which heavily restricts its performance. Though Planetoid combines supervised information with an unsupervised loss, there is information loss of graph structure during random sampling. To avoid that problem, GCN and GAT employ the neighborhood aggregation scheme to boost performance. GAT outperforms GCN as it can model different relations to different neighbors rather than with a pre-defined order. DGCN further jointly models both local and global consistency, yet its global consistency is still obtained through random walks. As a result, the information in the graph structure might lose in DGCN as well. On the contrary, the proposed H-GCN manages to capture global information through different levels of convolutional layers and achieves the best results among all four datasets.

Besides, on the NELL dataset, there are fewer training samples per class than in citation networks. Under such circumstance, training nodes are further away from testing nodes on average. The baseline models with the restricted receptive field are unable to propagate the features and the supervised information of the training nodes to other nodes sufficiently. As a result, the proposed H-GCN with increased receptive fields and deeper layers obtains more promising improvements than baselines.

\subsection{Impact of Scale of Training Data}

We suppose that a larger receptive field in the convolutional model promotes the propagation of features and labels on graphs. To verify the proposed H-GCN can get a larger receptive field, we reduce the number of training samples to check if H-GCN still performs well when limited labeled data is given. As in nature, there are plenty of unlabeled data; it is also of great significance to train the model with limited labeled data. In this section, we conduct experiments with different numbers of labeled instances on the Pubmed dataset. We vary the number of labeled nodes from 20 to 5 per class, where the labeled data is randomly chosen from the original training set. All parameters are the same as previously described. The corresponding performance in terms of accuracy is reported in Table \ref{tab:labeled-data}.

\begin{table}
	\centering
	\small
    \begin{tabular}{ccccc}
		\toprule
		Method & 20    & 15    & 10    & 5 \\ \midrule
		GCN   & 79.0\% & 76.9\% & 72.2\% & 69.0\% \\
		GAT   & 79.0\% & 77.3\% & 75.4\% & 70.3\% \\
		DGCN  & 79.3\% & 77.4\% & 76.7\% & 70.1\% \\
		H-GCN & {\bf 79.8\%} & {\bf 79.3\%} & {\bf 78.6\%} & {\bf 76.5\%} \\
    	\bottomrule
    \end{tabular}
	\caption{Results of node classification in terms of accuracy on Pubmed with labeled vertices varying from 20 per class to 5.}
	\label{tab:labeled-data}
\end{table}

From the table, it can be observed that our method outperform other baselines in all cases. With the number of labeled data decreasing, our method obtains a more considerable margin over these baseline algorithms. Especially when only five labeled nodes per class ($\approx$ 0.08\% labeling rate) are given, the accuracy of H-GCN exceeds GCN, DGCN, and GAT by 7.5\%, 6.4\%, and 6.2\% respectively. When the number of training data decreases, it is more likely for an unlabeled node to be further away from these labeled nodes. Only when the receptive field is large enough can information from those training nodes be captured. As the receptive field of GCN and GAT does not exceed 2-hop neighborhoods, supervised information contained in the training nodes cannot propagate sufficiently to other nodes. Therefore, these baselines downgrade considerably. However, owing to its larger receptive field, the performance of H-GCN declines slightly when labeled data decreases dramatically. Overall, it is verified that the proposed H-GCN with increased receptive fields is well-suited when training data is extremely scarce and thereby is of significant practical values.

\subsection{Ablation Study}
To verify the effectiveness of the proposed coarsening and refining layers, we conduct ablation study on coarsening and refining layers and node weight embeddings respectively in this section. The results are shown in Table \ref{tab:ablation-study}.

\begin{table}
	\centering
	\resizebox{\columnwidth}{!}{
	\begin{tabular}{ccccc}
		\toprule
		Method & Cora  & Citeseer & Pubmed & NELL \\ \midrule
		\makecell{H-GCN without coarsen-\\ing and refining layers} & 80.3\% & 70.5\% & 76.8\% & 75.9\% \\ \midrule
		\makecell{H-GCN without node\\ weight embeddings} & 84.2\% & 72.4\% & 79.5\% & 79.6\% \\ \midrule
		H-GCN & 84.5\% & 72.8\% & 79.8\% & 80.1\% \\
		\bottomrule
	\end{tabular}
	}
	\caption{Results of the ablation study}
	\label{tab:ablation-study}
\end{table}

\paragraph{Coarsening and refining layers.}
We remove all coarsening and refining operations of H-GCN and compare its performance with the original H-GCN. Different from simply adding too many GCN layers, we preserve the short-cut connection between the symmetric layers in the ablation study. From the results, it is evident that the proposed H-GCN has better performance compared to H-GCN without coarsening mechanisms on all datasets. It can be verified that the coarsening and refining mechanisms contribute to the performance improvements since they can obtain global information with larger receptive fields.

\paragraph{Node weight embeddings.}
To study the impact of node weight embeddings, we compare H-GCN with no node weight embeddings used. It can be seen from results that the model with node weight embeddings performs better, which verifies the necessity to add this embedding vector in the node embeddings.

\subsection{Sensitivity Analysis}

Last, we analyze hyper-parameter sensitivity. Specifically, we investigate how different numbers of coarsening layers and different numbers of channels will affect the results respectively. The performance is reported in terms of accuracy on all four datasets. While one parameter studied in the sensitivity analysis is changed, other hyper-parameters remain the same.

\begin{figure}
	\centering
	\subfloat[Coarsening layers]{
		\label{fig:coarsening-refining-layers}
		\includegraphics[width=0.495\columnwidth]{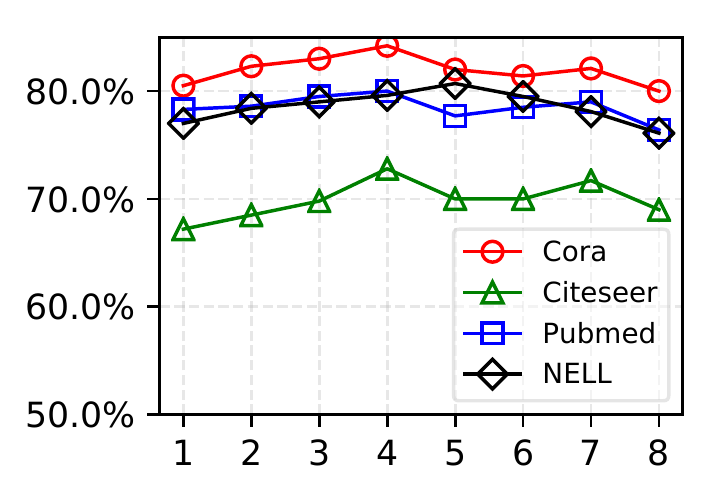}
	}
	\subfloat[Channel numbers]{
		\label{fig:channel-numbers}
		\includegraphics[width=0.495\columnwidth]{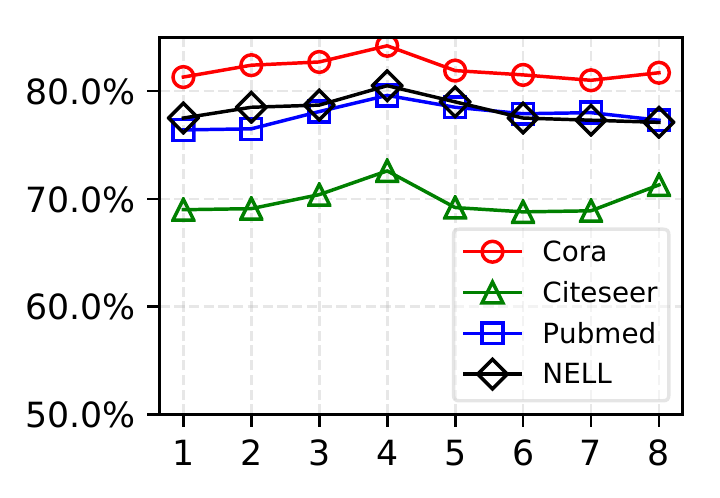}
	}\\
	\caption{Results of H-GCN with varying layers and channels in terms of accuracy.}
\end{figure}

\paragraph{Effects of coarsening layers.}
Since the coarsening layers in our model control the granularity of the receptive field enlargement, we experiment with one to eight coarsening and symmetric refining layers, where the results are shown in Figure \subref*{fig:coarsening-refining-layers}. It is seen that the performance of H-GCN achieves the best when there are four coarsening layers on three citation networks and five on the knowledge graph. It is suspected that, since less labeled nodes are supplied on NELL than others, deeper layers and larger receptive fields are needed. However, when adding too many coarsening layers, the performance drops due to overfitting.

\paragraph{Effects of channel numbers.}
Next, we investigate the impact of different amounts of channels on the performance. Multiple channels benefit the graph convolutional network model, since they explore different feature subspaces, as shown in Figure \subref*{fig:channel-numbers}. From the figure, it can be found that the performance improves with the number of channels increasing until four channels, which demonstrates that more channels help capture accurate node features. Nevertheless, too many channels will inevitably introduce redundant parameters to the model, leading to overfitting as well.

\section{Conclusion}

In this paper, we have proposed a novel hierarchical graph convolutional networks for the semi-supervised node classification task. The H-GCN model consists of coarsening layers and symmetric refining layers. By grouping structurally similar nodes to hyper-nodes, our model can get a larger receptive field and enable sufficient information propagation. Compared with other previous work, our proposed H-GCN is deeper and can fully utilize both local and global information. Comprehensive experiments have confirmed that the proposed method consistently outperformed other state-of-the-art methods. In particular, our method has achieved substantial gains over them in the case that labeled data is extremely scarce.

\section*{Acknowledgements}

This work is jointly supported by National Natural Science Foundation of China (61772528) and National Key Research and Development Program (2016YFB1001000, 2018YFB1402600).

\bibliographystyle{named}
\bibliography{ijcai19}

\end{document}